\documentclass{article}

\usepackage{amssymb,amsmath,mathtools,xcolor,graphicx,xspace,colortbl,ragged2e,rotating} %
\usepackage{amsmath}  
\usepackage{boxedminipage}  
\usepackage{float}  
\usepackage{ragged2e}  
\usepackage{xcolor}  
\graphicspath{{teleological_castagnoli march 16 2025_graphics/}{teleological_castagnoli march 16 2025_tcache/}{teleological_castagnoli march 16 2025_gcache/}}
\DeclareGraphicsExtensions{.pdf,.eps,.ps,.png,.jpg,.jpeg}

\begin{document}
\title{                                                                                                                                                                                                                        The physical basis of teleological evolutions 
}
\author{G. Castagnoli\\$$Formerly: Elsag Bailey ICT Division and Quantum Laboratory\protect\footnote{
giuseppe.castagnoli@gmail.com
}
}
\maketitle
\begin{abstract}We show that the sheer existence of a quantum computational speedup logically implies the mutually exclusive $or$ of well-defined causal loops. In each of them, it is as if the problem-solver knew in advance one of the possible halves of the information about the solution she will produce and measure in the future and could use this knowledge to produce the solution with fewer computation steps. Involving only the measurements of commuting observables, quantum algorithms are submitted to the same (classical) logic and should therefore host the quantum superposition of the causal loops in question. However, their ordinary quantum description does not and, being causal in character, cannot describe causal loops. It must therefore be incomplete and is in fact completed by time-symmetrizing it. This leaves the unitary part of the ordinary quantum description of the quantum algorithm mathematically unaltered but changes the behavior of causality along it. The single causal process of the ordinary quantum superposition is replaced by the quantum superposition of the causal loops in question. In their completed quantum description, quantum algorithms respond to the pre-scientific notion of teleological evolution, that is, an evolution toward a goal (the solution) with an attractor in the goal it will produce in the future (the solution again). Once applied to the evolutions of the living, this notion was dismissed with the advent of modern science for the alleged absence of a physical basis. We show that, under a quantum cosmological interpretation of the Anthropic Principle, the same quantum superposition of causal loops underlying the teleological character of quantum algorithms becomes the missing physical basis of the teleological character of natural evolutions.
\end{abstract}

\section{Introduction
}
This work proposes a physical basis for the notion of teleological evolution, that is, an evolution toward a goal with an attractor in the very goal it will produce in the future. It is, of course, a puzzling notion since it implies a causal loop. For over two millennia, it has been applied by natural philosophy first to the functional structure of the living (Aristotle: teeth were created for the purpose of chewing) and then, with the advent of the notion of evolution of the living, to the latter. It was then dismissed with the advent of modern science because of its alleged lack of physical basis. Of course, this is because the teleological notion implies retrocausality, even today a very controversial if not ostracized notion -- certainly a notion never accepted by mainstram science.

What we think should change this (centuries-old) position of science is the advent of quantum mechanics and, more specifically, the relatively recent discovery of the quantum computational speedup. In fact, we will show that quantum algorithms are scientific examples of teleological evolutions: it is their teleological character that yields their quantum computational speedup over the corresponding classical algorithms.

  Also with coauthors, we have developed the retrocausal explanation of the quantum speedup $\left [1 -5\right ]$. In it, it is as if the problem-solver knew in advance, before beginning her problem-solving action, one of the possible halves of the information about the solution of the problem she will produce and measure in the future and could use this knowledge to produce the solution with fewer computation steps (oracle queries) than the minimum required in the classical case. The number of computation steps needed in the optimal quantum case is that needed to solve the problem in an optimal logical way benefiting of the advanced knowledge in question. By the way, more precisely, what the problem-solver knows in advance is the mutually exclusive $or$ of all the possible halves of the information about the solution.

In the previous works, we showed that that explanation fits all the major quantum algorithms, see $\left [4\right ]$. In the present work, we provide an elementary derivation of it, truly elementary and perfectly checkable, one just had to think about it. We analize Grover's quantum search algorithm $\left [6\right ]$ from the standpoint of classical logic (we then generalize to oracle quantum computing). This is legitimate: involving only the measurements of commuting observables (the initial measurement of the setting of the problem and the final measurement of the corresponding solution), quantum algorithms are submitted to classical logic. In a few lines, we show that the sheer existence of the quantum speedup of Grover algorithm logically implies the mutually exclusive $or$ of the above mentioned causal loops.

Of course, this retrocausal explanation of the quantum speedup is a unifying quantitative one. Given an oracle problem (any one), it provides the number of computation steps (oracles queries) needed to solve it in an optimal quantum way; it is the number needed to solve it in an optimal logical way benefiting of the advanced knowledge of one of the possible halves of the information about the solution. Let us call this \textit{the advanced knowledge rule}. 

Note that, despite its obvious convenience, this rule is absent from mainstream quantum information, which relies on the notion of \textit{causal evolution}, i.e. an evolution where causality always goes forward in time. Here, the only way of knowing the number of computation steps (oracle queries) needed to solve an oracle problem in an optimal quantum way is finding the quantum algorithms that does it. Of course, this is an object of high ingenuity, very uncertain achievement, and different from oracle problem to oracle problem. As far as we know, excluding extremely simple quantum algorithms, this object has been reached in only one case, that of Grover's quantum search algorithm.

There is, however, a catch. To date, no work involving retrocausality, although published in mainstream physics journals, sometimes by famous authors, has ever made it into mainstream science.

There is also the opposite ``however''. However, dealing with the quantum speedup, the retrocausality we are dealing with presents a substantial novelty with respect to all the forms proposed before. Indeed,  it has two very significant pro's:

\medskip

(a) One is that it is logically implied by the sheer existence of the quantum speedup in an elementary and perfectly checkable way (as we will soon see).

\medskip

(b) The other is that it turns out to be just an \textit{interpretation} of the unitary interplay between quantum superpositions and interferences that characterizes the quantum computation of the solution of a problem. In other words, suitable unitary interplays between quantum superpositions and quantum interferences -- the two pillars of quantum mechanics -- generate retrocausal effects. Retrocausality is implicit in mainstream quantum mechanics!

\medskip

Note that, despite being just an interpretation of the current quantum description of quantum algorithms, the retrocausality we are dealing with is a retrocausality in the fullest sense of the world. Knowing in advance half of the information about a future measurement outcome means information sent backward in time from the future to the past, full fledged retrocausality indeed.

We would argue that the two above pro's should guarantee to retrocausality, at last, right of citizenship in mainstream quantum mechanics. We believe that this is the main result of the present work. Let us have have, anyway, a summary of the rest. We will see that everything fits together.

We now move from the logic of quantum algorithms to their physics. 

Of course, their current quantum description, a causal one (i.e. with causality always going forward in time), does not and cannot describe causal loops. We must conclude that it is incomplete, of course in a fundamental way. However, first, we should get rid of a trivial incompleteness also present in the current quantum description of quantum algorithms. Being limited to the unitary computation of the solution (where the quantum speedup resides) and the final measurement of the latter, their current quantum description is trivially incomplete for the lack of the initial measurement. This is also trivially emended, by adding to their current quantum description that of the upstream process of setting the problem. An initial measurement in a quantum superposition (or, indifferently, mixture) of all the possible problem-settings selects one of them at random. We call this extension of the current quantum description of the quantum algorithm its \textit{ordinary} \textit{quantum description}, ``\textit{}ordinary'' because it is still the causal quantum description of mainstream science.

We will show that the only possible way of completing it, by making it to describe the causal loops logically implied by the sheer existence of the quantum speedup, is by time-symmetrizing it. This leaves the unitary part of the ordinary quantum description of the quantum algorithm mathematically unaltered by changes the behavior of causality along it. The single causal process of the ordinary quantum description is replaced by the quantum superposition of the causal loops whose mutually exclusive $or$ is logically implied by the sheer existence of the quantum speedup.

This completion of the ordinary quantum description of quantum algorithms immediately generalizes to that of quantum correlation (in quantum algorithms, between the setting and the solution of the problem). More specifically, it extends to any unitary evolution between two one-to-one correlated measurement outcomes that does not change the eigenvalue of the observable selected by the initial measurement (e. g., the problem-setting is not changed by the unitary computation of the solution).

By the way, this is also the case of quantum nonlocality, that is, the measurement (in the same basis) of two maximally entangled observables in a state of spatial separation between the two respective subsystems. Time-symmetrizing its ordinary quantum description, exactly as we did with that of quantum algorithms, yields a quantum superposition of causal loops that play the role of the hidden variables envisaged by Einstein and the others in their celebrated EPR paper $\left [7\right ]$. They locally tell the not yet measured observable the state to assume in view of its future measurement.

All the above also says that quantum algorithms are full fledged teleological evolutions, indeed, evolutions toward a goal, the solution of the problem, with an attractor in the very goal they will produce in the future, the solution of the problem again.

Of course, showing that quantum algorithms are scientifically justified teleological evolutions raises the question whether this should change the position of science on natural teleological evolutions.

It should be noted that whether to abandon the teleological view of the functional structures and the evolutions of the living has long been debated among evolutionary biologists and is sometimes discussed also today. This is because it is a view that fits the living things in a very natural way. Therefore, we believe that checking whether the present scientific notion of teleological evolution is applicable also to the natural sciences should be a real must.

What we found would seem to be significant. It is that a quantum cosmological interpretation of the Fine-tuned Universe version of the Anthropic Principle, we would say ``very surprisingly'', yields that the evolution of the universe (under the fundamental physical laws and constants) toward the ``value of life'' (either ``existence of sentient life'' of ``absence of life'') is formally identical to a quantum search algorithm. In either case, the evolution could occur with quantum speedup and thus be teleological in character; in either case, it should occur with quantum speedup even if the search is random. Of course, the ``quantum cosmological assumption'' $\left [8\right ]$ is indeed an assumption, that is, is completely hypothetical in character. However, the long-standing claim that the teleological notion is devoid of any possible physical basis would in any case be falsified: indeed, a possible physical basis does exist.

\section{
Rationale}
It should already be clear from the introduction that this work is unconventional. In effect, it is a different way of approaching quantum information. Therefore, before going to the formalizations and further detail, in the present ``Rationale'' we provide a conceptual presentation of the work.

The Rationale is divided into two parts. The first provides a physical basis to the teleological character of quantum algorithms and, more in general, the building up of quantum correlation (in quantum algorithms, between the setting and the solution of the problem). The second examines the possibility that the physical basis of the teleological character of quantum algorithms also applies to the evolutions of the living.

\subsection{
The physical basis of the teleological notion in the quantum domain}
Let us provide the reason why quantum algorithms would be teleological evolutions.

We must begin with an example of quantum speedup. The context, of course, is problem-solving. We consider the following very simple problem. Bob, the problem-setter, hides a ball in a chest of four drawers -- we call the number of the drawer with the ball the \textit{problem-setting}. Alice, the problem-solver, is to locate the ball by opening drawers. In the classical case, she has to open up to three drawers. If the ball is only in the third drawer opened, she has solved the problem; if it is not even in that drawer, it must be in the only drawer not yet opened and she has solved the problem of locating the ball as well. In the optimal quantum case -- by the optimal quantum search algorithm of Grover $\left [6\right ]$ -- she always solves the problem by opening just one drawer.  This would obviously be impossible in the classical case. 

Let us show, by the way very simply, that always solving the problem of locating the ball by opening only one drawer logically implies the mutually exclusive $or$ of well defined causal loops. 

Here, by ``opening only one drawer'', we mean performing only one oracle query. Of course, only one drawer is opened but for a quantum superposition of all the possible drawer numbers. In the following, we analyze this behavior of the quantum algorithm from the standpoint of logic, classical logic of course. Indeed, involving only the measurements of commuting observables (the problem-setting and the corresponding solution of the problem) quantum algorithms are submitted to classical logic (there is no special ``quantum logic'' involved here). 

First, the fact of always solving the problem by opening just one drawer is logically equivalent to the problem-solver always knowing in advance, before she begins her problem-solving action, that the ball is in one of two given drawers and benefiting of this knowledge to solve the problem by opening just one drawer. Indeed, if the ball is in the drawer opened, she has solved the problem; if not, it must be in the other drawer and she has solved the problem of locating the ball as well, in either case having opened just one drawer. Of course, what Alice should know in advance is the mutually exclusive $or$ of the three possible ways of pairing the number of the drawer with the ball with that of another drawer. Note that this also means knowing in advance half of the information that specifies the solution of the problem she will produce and measure in the future.

Then the question becomes: from where does Alice, before beginning her problem-solving action, always get that information about the solution? Since, to her, the only possible source of it is the solution of the problem she will produce and measure in the future, the only possible answer is that it comes to her from her future measurement of the solution. Of course, this implies retrocausality, more specifically  a causal loop.

The above immediately generalizes to any number, $N$, of drawers. Grover algorithm yields a quadratic quantum speedup, that is, always solves the problem by opening $\ensuremath{\operatorname*{O}}\left (\sqrt{N}\right )$ drawers against the $\ensuremath{\operatorname*{O}}\left (N\right )$ of the best classical case\protect\footnote{
The four drawer instance of Grover algorithm gives the solution with certainty; with more than four drawers, by ``Grover algorithm'' we will always mean Long's version of it $\left [9\right ]$, which always yields the solution with certainty.
}. Also now, this quantum speedup can be logically justified by the fact that Alice always knows in advance, in a mutually exclusive way, one of the possible halves of the information about the number of the drawer with the ball. This means that she knows in advance that the ball is (in a mutually exclusive way) in one of the possible tuples  of $\sqrt{N}$ drawers. She can then solve the problem in the best possible logical way by opening  $\ensuremath{\operatorname*{O}}\left (\sqrt{N}\right )$ drawers, which is indeed what Grover algorithm does.

We can conclude that the quantum speedup of Grover algorithm logically implies the mutually exclusive $or$ of well defined causal loops, each for each possible way of taking half of the information about the solution. In each of them, it is as if the problem-solver knew in advance one of the possible halves of the information about the solution she will produce and measure in the future; she then uses this knowledge in the best possible logical way to produce the solution by opening $\ensuremath{\operatorname*{O}}\left (\sqrt{N}\right )$ drawers.

Note the simplicity of the above argument. It is so simple that it runs the risk of being overlooked. Instead, we argue, it should be of fundamental importance. Whether causal loops are physical or unphysical is a controversial subject. Of course, they are never used in mainstream quantum mechanics. According to the argument in question, instead, they must be implicit in Grover quantum search algorithm. Moreover, as we shall see, they are necessarily implicit in any quantum algorithm that yields a quantum speedup over what is classically possible.

Let us now move from logic to physics, of course quantum physics (the quantum algorithm indeed). Naturally, the current quantum description of quantum algorithms does not describe any causal loop. Furthermore, being the customary ``causal'' quantum description of evolutions, where causality always goes in the same time-direction, it could not. Consistently with the present line of thought, we must conclude that this quantum description is incomplete.

This is, of course, a bold claim since, as it is, the quantum description has worked amazingly well for exactly a century to date. However, the boldness of the claim can be mitigated in two ways:

\medskip

(i) First, it is a claim with an illustrious precedent, although apparently forgotten by the scientific community. We mean  the 1935 argument of Einstein et al on the incompleteness of the quantum description $6$. We will see that the incompleteness of the quantum description of quantum algorithms is exactly the same lamented by Einstein et al for quantum nonlocality; correspondingly, the way to complete the quantum description we will provide for quantum algorithms, will also work for quantum nonlocality providing the hidden variables -- here the causal loops -- needed in both cases.

\medskip

(ii) Second, let us distinguish between the mathematical and the physical description of a quantum process. The completion of the quantum description we will propose leaves the unitary part of the quantum algorithm mathematically unchanged but changes the description of the behavior of causality along it. The single causal process of the ordinary quantum description is replaced by the quantum superposition of the causal loops whose mutual exclusive $or$ is logically implied by the sheer existence of the quantum speedup.

\medskip

We show how this result is reached. Here we provide conceptually what will be formalized in Section 3.. Obviously, to start with, we should choose the kind of quantum description of quantum algorithms to complete. This is said because it cannot be their usual quantum description, which is itself trivially incomplete. Indeed, it is limited to the unitary process of solving the problem, where the quantum speedup resides, and the final measurement of the solution. It thus lacks the initial measurement. Of course, the canonical quantum description of a quantum process must consist of initial measurement, unitary evolution, and final measurement.

Naturally, this problem is simply amended. It suffices to extend the quantum description of the process of solving the problem to the upstream process of setting the problem. An initial measurement randomly selects a problem-setting (e. g. the number of the drawer with the ball) out of a quantum superposition (or, indifferently, mixture) of all the possible problem-settings, the unitary computation of the corresponding solution and the final measurement of the latter follow; for simplicity, we will always assume that the computation of the solution yields it with certainty. In view of what will follow, we note that, as a consequence, the unitary transformation goes from the initial to the final measurement outcome despite comprising the final measurement; thus, in particular, the process connecting the two measurement outcomes is time-reversible.

Incidentally, this way of describing quantum algorithms, which highlights the quantum correlation between the problem-setting selected by the initial measurement and the corresponding solution selected (read) by the final measurement, will play a key role in the present work. We call it\textit{ the ordinary quantum description of quantum algorithms}; ordinary because it is still the customary causal quantum description of mainstream quantum mechanics that cannot describe causal loops.

We show now how to complete this ordinary quantum description in a way that describes the causal loops we are dealing with; we will later see that it is the only possible way.

Always for simplicity, we assume that the solution is an invertible function of the problem-setting\protect\footnote{
This constraint is easily removed. It suffices that, at the end, the problem solver first measures the problem-setting, which of course remains unaltered through the computation of the solution, then reads the solution.
}. We should proceed as follows.

We require that the determination of the problem-setting and thus that of the corresponding solution, instead of being entirely ascribed to the initial measurement of the problem-setting, is evenly shared between the initial measurement of the problem-setting and the final measurement of the corresponding solution. Since there is a plurality of (logically) mutually exclusive ways of evenly sharing, at the end we should take their quantum superposition. 

Note that, since the unitary evolution between the two measurement outcomes (i.e., the computation of the solution) never changes the problem-setting selected by the initial measurement, the latter measurement, or any part of it, can be postponed at will, thus also making it simultaneous with the final measurement. Of course, this allows the sharing of the selection in question but does not justify the fact of assuming it. As we shall see, the justification is that it is the only way to generate the causal loops the quantum algorithm must host.

In each time-symmetrization instance, the state selected by the initial measurement (not a single problem-setting but a suitable partition of the quantum superposition of all the problem-settings) unitarily propagates forward in time until becoming the state immediately before the final measurement (a quantum superposition of the problem-settings selected by the initial measurement each tensor product the corresponding solution). Time-symmetrically, the state selected by the final measurement (a single problem-setting tensor product the corresponding solution) propagates backward in time, by the inverse of the time-forward unitary transformation, until replacing the previous outcome of the initial measurement.

Of course, propagating backward in time the outcome of a measurement is unconventional. However, first, note that the outcome of the final measurement -- a single problem-setting tensor product the corresponding solution -- coincides with that of the ordinary quantum description of the quantum algorithm. As a consequence, mathematically (i.e. disregarding the time-direction of causality), its  backward in time propagation, by the inverse of the time-forward unitary transformation, is the unitary part of the ordinary quantum description of the quantum algorithm back again. This backward in time propagation just undoes the sharing of the selection between initial and final measurements; it is therefore legitimate. Second, as we shall see, it is mandatory to complete the quantum description of the quantum algorithm in a way that describes the causal loops that it must host.

We should eventually take the quantum superposition of all the time-symmetrization instances. 

We call the above sequence of operations the time-symmetrization of the ordinary quantum description of the quantum algorithm, because it is indeed the time-symmetrization of the time-reversible quantum process between the initial and final one-to-one correlated measurement outcomes.

We should note what follows:

\medskip

(i) each time symmetrization instance consists of a time-forward propagation (by a unitary transformation) going from the outcome of the initial measurement to the outcome of the final measurement and a time-backward propagation (by the inverse of the previous unitary transformation) going from the outcome of the latter measurement until replacing the previous outcome of the initial measurement. One can see that this is a non-trivial causal loop, namely one that changes its own past.

\medskip

(ii) the  backward in time propagation of each time-symmetrization instance, which inherits both selections, is an instance of the time-symmetrized  quantum algorithm. Since all backward in time propagations are the same, any one of them is the time-symmetrized quantum algorithm.

\medskip

Summing up, we can see how the time-symmetrization process we are dealing with completes the ordinary (causal) quantum description of the quantum algorithm. First, it leaves the unitary part of the ordinary quantum description mathematically unaltered. Second, it changes the description of the behavior of causality along that part: the single causal process of the ordinary quantum description is replaced by a quantum superposition of causal loops.

We will see that, as viewed by the problem solver, the causal loops in quantum superposition we are dealing with are exactly the causal loops whose mutual exclusive $or$ is logically implied by the sheer existence of the quantum speedup.

Let us explain the reason for the specification ``as viewed by the problem-solver''. The ordinary quantum description of the quantum algorithm is with respect to the customary observer external to the quantum process. It cannot be with respect to the problem-solver: it would tell her, before she begins her problem-solving action, the problem-setting selected by the initial measurement and thus the solution of the problem (here, they are both the number of the drawer with the ball). To have the quantum description with respect to the problem-solver (\textit{relativized} to her), we should conceal from her the problem-setting selected by the initial measurement (see Subsection 3.1.2.)

Similarly, we should relativize to the problem-solver the causal loops generated by time-symmetrizing the ordinary quantum description of the quantum algorithm, by hiding from her the part of the problem-setting selected by the initial measurement (see Subsection3.1.4).

Also note that the time-symmetrization we are dealing with is the only possible way of completing the ordinary quantum description of the quantum algorithm in a way that describes the causal loops logically implied by the sheer existence of its quantum speedup. In fact, with the kind of process we are dealing with, to describe a causal loop (any one), we must properly share the selection of the problem-setting between initial and final measurements, with the part of the selection performed by the initial measurement propagating forward in time and that performed by the final measurement backward in time. To obtain exactly the causal loops whose mutually exclusive $or$ is logically implied by the sheer existence of the quantum speedup, this sharing must be even. This exactly identifies the time-symmetrization we have performed. It also shows that this time-symmetrization is mandatory: it is the only way of generating the quantum superposition of the causal loops whose mutually exclusive $or$ is logically implied by the sheer existence of the quantum speedup and that the quantum algorithm must host.

Further below we will show that all the above holds for a large class of quantum algorithms, those solving oracle problems. Note that, so completed, the quantum description of quantum algorithms says much more than their ordinary quantum description; in fact, it highlights the following unifying quantitative feature of the quantum speedups:

\medskip

\textit{Given an oracle problem, the number of computation steps (oracle queries) needed to solve it in an optimal quantum way is that needed to solve it in an optimal logical way benefiting of the advanced knowledge of half of the information that specifies the solution among all the possible solutions.}

\medskip

As is well known, in mainstream quantum computing science (which naturally relies on the notion of causal evolution), there is no unified quantitative description of the different quantum speedups. The only way of knowing the number of computation steps needed to solve a given oracle problem in an optimal quantum way is finding an optimal quantum algorithm that does that, naturally an object of great ingenuity and uncertain achievement.

Let us call the above statement in italics \textit{the advanced knowledge rule}. This rule had already been achieved in the retrocausal approach to the quantum speedup of ourselves and coauthors $\left [1 -5\right ]$. In $\left [4\right ]$, we had shown that it quantitatively explains the quantum speedup of all the major quantum algorithms. However, in these works, the time-symmetrization of the ordinary quantum description of the quantum algorithm was justified only by the fact that it turns out to produce, indeed, a unified quantitative description of the quantum speedups. 

Now, this time-symmetrization has been shown to be necessary. It is the necessary consequence of the fact -- we think inescapable -- that the sheer existence of a quantum speedup over what is classically possible logically implies the mutual exclusive $or$ of well defined causal loops that must be implicit in the quantum algorithm. This should be a significant advancement. Causal loops, hitherto foreign to mainstream quantum  mechanics, would instead be common in it, at least as common as quantum algorithms are.

We open a parenthesis to explain why we provided a review of a retrocausal approach to the quantum speedup that has already been published. The point is that the approach in question is probably little known as it never entered mainstream physics. The latter is a fate common to most publications dealing with retrocausality, starting with the famous1949 one by Wheeler and Feynman $\left [10\right ]$. Although, at the time, the latter gave rise to several discussions, none of it ever entered mainstream physics. Of course, the idea that the effect can precede the cause can be difficult to accept. However, we have already seen in the Introduction, that this effect preceding the cause must be implicit in quantum mechanics as it can always be replaced by a suitable, unitary, interplay between quantum superpositions and interferences, of course the two the pillars of quantum mechanics.

Back to our previous line of thought, in Subsection 3.2, we will see that the need of completing the quantum description by time-symmetrizing it is not limited to quantum algorithms. It extends as follows:

\medskip

\textit{The ordinary quantum description of any unitary evolution between an initial and final one-to-one correlated measurement outcomes that does not change the eigenvalue selected by the initial measurement is incomplete and is completed by time-symmetrizing it. }

\medskip

 The rule in question extends to quantum correlation the need of completing its ordinary quantum description by time symmetrizing it, what of course generates a quantum superposition of causal loops. Let us call it \textit{the causal loops rule of quantum correlation}.

In the first place, this rule says that the unitary building up of one-to-one quantum correlation between two initially unrelated observables that does not change the eigenvalue of the observable selected by the initial measurement can always occur with quantum speedup (see Subsection 3.2 for further detail). Note that, if the process is not a quantum algorithm, the notion of advanced knowledge on the part of the problem-solver of half of the information about the solution must be replaced by that of a corresponding reduction of the dimension of the Hilbert space in which the search for one-to-one correlation between the two observables takes place.

In the second place, the rule also applies to the case that the unitary evolution does not build up any quantum correlation between the two observables because they are maximally entangled to start with. This is the case of quantum nonlocality, where there is one-to-one quantum correlation between the outcomes of measuring (in the same basis) two maximally entangled observables in a state of spatial separation between the two respective subsystems. Of course, the eigenvalue selected by either measurement is not changed by the unitary process of spatial separation between the two subsystems. 

As shown in $\left [5\right ]$, the causal loops generated by time-symmetrizing the ordinary quantum description of quantum nonlocality locally go from the initial to the final measurement outcome via the time the two subsystems were not yet space-separate, where they locally tell the not yet measured observable the state to assume in view of its future measurement. Therefore, these causal loops exactly play the role of the hidden variables that Einstein and the others $\left [7\right ]$ envisaged to locally explain quantum nonlocality.

The above also says that the incompleteness that we are presently ascribing to the ordinary quantum description of quantum algorithms is exactly the same that Einstein and the others ascribed to that of quantum nonlocality. Of course, in both cases we are dealing with the incompleteness of the quantum description of a quantum correlation. 

By the way, we do not know why the diagnosis of the incompleteness of the quantum description made by Einstein and the others has been removed by the scientific community despite Einstein's authority. Perhaps, because a satisfactory completion of it has never been provided. Of course, the price to pay for the present way of completing it is opening to retrocausality, a feature still avoided by mainstream quantum mechanics.

With this, we end a first part of the Rationale.

\subsection{
Does the physical basis of the teleological character of quantum algorithms also apply to natural evolutions?
}
As we have seen, completing the ordinary quantum description of quantum algorithms highlights their teleological character, that is, their being evolutions toward a goal (the solution) with an attractor in the very goal they will produce in the future (the solution again). We can say that, in the quantum world, we have found the missing physical basis for the notion of teleological evolution.

This, naturally, raises the question whether the physical basis in question also applies to the evolutions of the living to which the teleological notion, before the scientific revolution, was applied. The second part of the Rationale addresses this question. 

Incidentally, being thought without physical basis, and specially because of its finalistic character, the teleological view has been classified among the beliefs. This, however, should not conceal the fact that, to start with, it was motivated by obvious empirical reasons. Indeed, the teleological view fits the functional structures and the evolutions of the living much better than the causal view. Of course, besides being why it was conceived in the first place, this is why, in relatively recent times, whether to abandon it has long been debated among biologists (see Subsection 4.4.).

The question of whether the physical basis of the teleological character of quantum algorithms also applies to natural evolutions immediately raises a seemingly insurmountable problem. Quantum algorithms belong to the microscopic world; their implementations deal with a rather limited number of quantum bits and durations of a few tens microseconds. As things are now, larger sizes or durations soon bring to quantum decoherence and the process becomes classical. The evolutions of the living, instead, can be as large as the Earth and have durations of billions of years. They are therefore among the furthest things from the quantum processes of the microscopic world to which quantum algorithms belong. 

Of course, we are not discussing whether the chlorophyll cycle or orientation in migratory birds is quantum, which, as is well known, is considered to be a possibility. Here, the very enormity of the distance between the world of quantum algorithms and that of natural evolutions forbids any attempt to scale up toward the macroscopic world the quantum features of the microscopic one.

However, the very enormity of this distance suggests a possible solution. It is to adopt quantum cosmology $\left [8\right ]$, namely the assumption that the universe -- a perfectly isolated system since by definition there is nothing outside it -- evolves in a unitary quantum way. We could then think of scaling the quantum features down from the largest and longest-lasting evolution in existence, that of the universe. 

Let us show how a consequent quantum interpretation of the Anthropic Principle $\left [11 -14\right ]$, the Fine-tuned Universe version of it, could provide a solution to our problem.

As is well known, this version of the Anthropic Principle is a way of expressing the noticeable fact that the slightest change of the value of the fundamental physical constants since the beginning of time would have produced a radically different universe most probably unable to accommodate life.

This version of the principle just asserts that there is a correlation (we assume for simplicity one-to-one) between the value (for values) of the fundamental constants since the beginning of time and the ``value of life'' today (in particular, ``existence of sentient life'' or ``absence of life''). Of course, this correlation must have been built up by the evolution of the universe under the fundamental laws and constants. In fact, at the beginning of time, the value of the fundamental constants and that of life were unrelated: no matter the value of the fundamental constants, there could have been no life.

Under the assumption that the evolution of the universe is unitary quantum, the correlation asserted by the anthropic principle becomes quantum, and we have seen that this kind of building up of quantum correlation (unitary and never changing then eigenvalue selected by the initial measurement) can always occur with quantum speedup (see Subsection 3.2 for further detail). Indeed, the unitary building up of quantum correlation between the value of the fundamental constants since the beginning of time and the value of life today is formally identical to a quantum search algorithm that unitarily builds up a quantum correlation between the value of the problem-setting since its beginning and the value of the solution at its end -- one can see that this is a way of viewing what quantum algorithms do.

The evolution of the universe toward the value of life can therefore occur with quantum speedup and consequently be teleological in character -- the ``can occur'' is said in the sense that there is certainly the room for its occurring with a quantum speedup\protect\footnote{
Note that, in problem-solving, advanced knowledge on the part of the problem-solver of half of the information about the solution does not imply a quantum speedup but means room for it; of course, in order that there is a quantum speedup, the quantum algorithm must benefit of this advanced knowledge.
}.

 In the universe we are in, what can occur with quantum speedup is the evolution of the universe toward life. This room for quantum speedup is then inherited by the evolution of life on Earth, even if decohered. In fact, first, evolution times must be the same for the evolution of life on Earth, obtainable in principle from that of the universe by tracing out all the rest, and its image in the evolution of the universe. Second, the resources employed by the former evolution cannot exceed those employed by an image that comprises them.

In the end, being able to occur with quantum speedup, the evolution of life on Earth can be teleological; the same would consequently hold for the evolutions of life, in particular those of species. Since it is reasonable, and conservative, to assume that the search for the value of life by the evolution of the universe under the fundamental physical laws and constants is random, its room for quantum speed should become an actual quantum speedup, as in quantum random search algorithms $\left [15\right ]$. 

 This ends the second part of the work. Of course, we should keep in mind that, being under the quantum cosmological assumption, this part is hypothetical.

Section 3. is about the teleological notion in the quantum framework; it is the formalization of the first part of the Rationale.

Section 4. is about the present, scientific, teleological notion in natural evolutions -- it is an extension and formalization of the second part of the Rationale.

Section 5. provides the prospects for further research; Section 6. tries a positioning of the work; Section 7. yields the conclusions.

\section{
The teleological notion in the quantum domain
}
This section provides the physical basis of the notion of teleological evolution in the quantum domain. It is in two parts. In the first, we formalize the notion of completing the ordinary quantum description of quantum algorithm by time-symmetrizing it. The second is the generalization from quantum algorithms to quantum correlation.

\subsection{
Completing the ordinary quantum description of quantum algorithms by time-symmetrizing it
}
The completion of the quantum description of quantum algorithms consists of the following steps:

\medskip

(1) Extend the usual quantum description of the quantum algorithm, limited to the process of solving the problem, to the upstream process of setting the problem. We will work on the simplest, four-drawer instance of Grover's quantum search algorithm and then generalize.

\medskip

(2) The ordinary quantum description of the quantum algorithm is with respect to the customary observer external to the quantum process. We show how to relativize it with respect to the problem solver, to whom the problem setting selected by the initial measure is to be kept hidden.

\medskip

\medskip (3) Complete the ordinary quantum description of the quantum algorithm by time-symmetrizing it.

\medskip

(4) Relativize with respect to the problem-solver the causal loops generated by time-symmetrizing the ordinary quantum description of the quantum algorithm, which are of course with respect to the customary external observer. Note that the causal loops whose mutually exclusive $or$ is logically implied by the sheer existence of the quantum speedup, each telling the problem-solver in advance one of the possible halves of the information about the solution, are of course with respect to her. 

\medskip

Steps (1) to (4) are developed in the following subsections.

\medskip

\subsubsection{Extending the usual quantum description of quantum algorithms to the process of setting the problem}
The usual quantum description of quantum algorithms is limited to the unitary process of solving the problem, where the quantum speedup resides, and the final measurement of the solution. It is trivially incomplete since it lacks the initial measurement. Of course, the canonical quantum description of a quantum process must consist of initial measurement, unitary evolution, and final measurement. We extend the quantum description of the process of solving the problem to the upstream process of setting the problem. An initial measurement of the problem-setting in a quantum superposition (or, indifferently, mixture) of all the possible problem-settings selects one of them at random. This is followed by the unitary evolution that computes the corresponding solution and the final measurement of the latter.

Let us introduce the notation. We make reference to the simplest, four drawer, instance of Grover algorithm:

\medskip

We number the four draws in binary notation: $00 ,01 ,10 ,11$.  

\medskip

We will need two quantum registers of two \textit{quantum bit}s each:

\medskip

A register $B$ under the control of the problem-setter Bob contains the \textit{problem-setting} (the number of the drawer in which the ball is hidden). The content of register $B$ is \textit{the problem-setting observable
$\hat{B}$   }of eigenstates and corresponding eigenvalues:

\medskip \medskip

$\vert 00 \rangle _{B}$ and $00$; $\vert 01 \rangle _{B}$ and $01$; $\vert 10 \rangle _{B}$ and $10$; $\vert 11 \rangle _{B}$ and $11$. 

\medskip 

              A register $A$ under the control of the problem-solver Alice contains the number of the drawer opened by her. The content of register $A$ is the observable
$\hat{A}$  of eigenstates and corresponding eigenvalues:

\medskip

$\vert 00 \rangle _{A}$ and $00$; $\vert 01 \rangle _{A}$ and $01$; $\vert 10 \rangle _{A}$ and $10$; $\vert 11 \rangle _{A}$ and $11$.

\medskip

Measuring
$\hat{A}$  at the end of the quantum algorithm yields the number of the drawer with the ball selected by the initial measurement, namely the solution of the problem. Note that
$\hat{B}$  and
$\hat{A}$  commute.

\medskip

The quantum description of the extended quantum algorithm, namely its \textit{ordinary quantum description} is: 

\medskip

\begin{equation}\begin{array}{ccc}\begin{array}{c}\;\text{time }t_{1}\text{, meas. of}\;\hat{B}\end{array} & t_{1} \rightarrow t_{2} & \text{time }t_{2}\text{, meas. of}\;\hat{A} \\
\, & \, & \, \\
\begin{array}{c}\left (\vert 00 \rangle _{B} +\vert  01 \rangle _{B} +\vert 10 \rangle _{B} +\vert  11 \rangle _{B}\right ) \\
\left (\vert 00 \rangle _{A} +\vert 01 \rangle _{A} +\vert 10 \rangle _{A} +\vert 11 \rangle _{A}\right )\end{array} & \, & \, \\
\Downarrow  & \, & \, \\
\vert 01 \rangle _{B}\left (\vert 00 \rangle _{A} +\vert 01 \rangle _{A} +\vert 10 \rangle _{A} +\vert 11 \rangle _{A}\right ) &  \Rightarrow \hat{\mathbf{U}}_{1 ,2} \Rightarrow  & \vert 01 \rangle _{B}\vert  01 \rangle _{A}\end{array}\tag{Table I}
\end{equation}

\medskip

Here and in the following we disregard normalization.

We should start from the quantum state in the upper-left corner of Table I diagram and then follow the vertical and horizontal arrows. Initially, at time $t_{1}$, both registers, each by itself, are in a quantum superposition of all their basis vectors. By measuring
$\hat{B}$  (the content of $B$) in this initial state, Bob selects the number of the drawer with the ball at random, say it comes out $01$ -- see the state of register $B$ under the vertical arrow\protect\footnote{
By the way, Bob would be free to unitarily change at will the sorted out number of the drawer with the ball. We omit this operation that would change nothing here.
}.

It is now Alice's turn. The unitary part of her problem-solving action -- by the unitary transformation $\hat{\mathbf{U}}_{1 ,2}$ -- produces the solution of the problem, of course, never changing the problem-setting. 

Note that we will never need to know $\hat{\mathbf{U}}_{1 ,2}$ (i.e. Grover algorithm). We only need to know that there can be a unitary transformation between the outcome of the initial measurement and that of the final measurement, what is always the case since the final measurement outcome is an invertible function of the initial one.

At the end of Alice's problem-solving action, at time $t_{2}$, register $A$ contains the solution of the problem, i.e. the number of the drawer with the ball -- bottom-right corner of the diagram. 

Eventually, by measuring
$\hat{A}$  (the content of $A$), which is already in one of its eigenstates, Alice acquires the solution with probability $1$ -- the quantum state in the bottom-right corner of the diagram naturally remains unaltered throughout this measurement. Note that, as a consequence, the process between the initial and the final measurement outcomes is unitary, and thus reversible, despite comprising the final measurement.

\subsubsection{
Relativizing the ordinary quantum description of the quantum algorithm with respect to the problem-solver
}

The ordinary quantum description of the quantum algorithm of Table I is with respect to the customary external observer, that is, an observer external to the quantum process. It is not a description for the problem-solver Alice. In fact, it would tell her, immediately after the initial measurement of
$\hat{B}$, the problem-setting selected by the initial measurement and thus the solution of the problem. By the way, here they are both the number of the drawer with the ball. Of course, in the quantum description of the quantum algorithm with respect to Alice, she must be concealed from the outcome of the initial measurement.

Since Alice's problem-solving action never changes the number of the drawer with the ball selected by the initial measurement and the observables
$\hat{B}$ and
$\hat{A}$  commute, describing this concealment is simple. It suffices to postpone   after the end of Alice's problem-solving action the projection of the quantum state associated with the initial measurement of
$\hat{B}$  (or, indifferently, to postpone the very measurement of
$\hat{B}$). The result is the following quantum description of the quantum algorithm with respect to Alice:

\medskip

\begin{equation}\begin{array}{ccc}\begin{array}{c}\text{time}\text{ }t_{1}\text{, meas. of}\;\hat{B}\end{array} & t_{1} \rightarrow t_{2} & \text{\text{time }\text{}}t_{2}\text{, meas. of}\;\hat{A} \\
\, & \, & \, \\
\begin{array}{c}\left (\vert 00 \rangle _{B} +\vert  01 \rangle _{B} +\vert 10 \rangle _{B} +\vert  11 \rangle _{B}\right ) \\
\left (\vert 00 \rangle _{A} +\vert 01 \rangle _{A} +\vert 10 \rangle _{A} +\vert 11 \rangle _{A}\right )\end{array} &  \Rightarrow \hat{\mathbf{U}}_{1 ,2} \Rightarrow  & \begin{array}{c}\vert 00 \rangle _{B}\vert  00 \rangle _{A} +\vert 01 \rangle _{B}\vert  01 \rangle _{A} + \\
\vert 10 \rangle _{B}\vert  10 \rangle _{A} +\vert 11 \rangle _{B}\vert  11 \rangle _{A}\end{array} \\
\, & \, & \Downarrow  \\
\, & \, & \vert 01 \rangle _{B}\vert  01 \rangle _{A}\end{array}\tag{Table II}
\end{equation}

\medskip 

 To Alice, the initial measurement of \textbf{$\hat{B}$} leaves the initial state of register $B$ unaltered. At the beginning of her problem-solving action she is completely ignorant of the number of the drawer with the ball selected by Bob -- see the state of register $B$ on the left of $ \Rightarrow \hat{\mathbf{U}}_{1 ,2} \Rightarrow $.

 The unitary transformation   $\hat{\mathbf{U}}_{1 ,2}$ is then performed for a quantum superposition of the four possible numbers of the drawer with the ball. The state at the end of it -- on the right of the horizontal arrows -- is a quantum superposition of tensor products, each a possible number of the drawer with the ball in register $B$ tensor product the corresponding solution (that same number) in register $A$. Since
$\hat{B}$ and
$\hat{A}$  commute, the final measurement of
$\hat{A}$  must project this superposition on the tensor product of the number of the drawer with the ball initially selected by Bob and the corresponding solution -- bottom-right corner of Table II diagram.

Note that this relativization relies on relational quantum mechanics $\left [16 ,17\right ]$ where the quantum state is relative to the observer (also note that, here, one observer is external and the other internal to the quantum process). In the following, we will continue to call the quantum description with respect to the external observer \textit{the ordinary quantum description of the quantum algorithm}. Instead we will always specify that the other quantum description is that with respect to the problem-solver.

\subsubsection{
Time-symmetrizing the ordinary quantum description of the quantum algorithm
}
As already said, in order to obtain the description of the quantum superposition of the causal loops whose mutually exclusive $or$ is implied by the sheer existence of an optimal quantum speedup, we should first time-symmetrize the ordinary quantum description of the quantum algorithm, which is with respect to the customary external observer, then relativize the causal loops obtained with respect to the problem-solver.

To start with, as anticipated in the Rationale, we should require that the selection of the information that specifies the sorted out pair of correlated measurement outcomes among all the possible pairs evenly shares between the initial and final measurements. The half information selected by the initial measurement should propagate forward in time, by  $\hat{\mathbf{U}}_{1 ,2}$ until becoming the state immediately before the final measurement. Time-symmetrically, the complementary selection performed by the final measurement should propagate backward in time, by
$\hat{\mathbf{U}}_{1 ,2}^{\dag }$, until replacing the previous outcome of the initial measurement.

 Note that the selection, by the initial measurement, of half of the information that specifies the problem-setting is made possible by two facts: (i) the measurement occurs in a quantum superposition of all the possible problem-setting, therefore it is sufficient to reduce it to the measurement of one of the possible halves of the bits of the problem-setting in question (in register $B$) and (ii) the problem-setting never changes along the unitary evolution between the two measurement outcomes, therefore the measurement of the other half of the bits can be postponed along that evolution until making it simultaneous with the final measurement.  Of course, as the even sharing of the selection can be performed in many possible mutually exclusive ways, eventually we should take their quantum superposition.

We consider the time-symmetrization instance where the initial measurement of
$\hat{B}$  reduces to the measurement of
$\hat{B}_{l}$, the left bit of the two-bit number contained in register $B$; say that it randomly selects left bit $0$. The final measurement of
$\hat{A}$  should correspondingly reduce to that of
$\hat{A_{r}}$  the right bit of the two-bit number contained in register $A$; say that it randomly selects right bit $1$. The randomly selected number of the drawer with the ball is thus $01$.

In this particular time-symmetrization instance, the ordinary quantum description of the quantum algorithm with respect to the external observer
of Table I becomes:

\medskip

\begin{equation}\begin{array}{ccc}\begin{array}{c}\;\text{time }t_{1}\text{, }\text{}\text{meas. of}\;\hat{B}_{l}\end{array} & t_{1 \rightleftarrows }t_{2} & \text{time }\text{}t_{2}\text{, meas. of}\;\hat{A_{r}} \\
\, & \, & \, \\
\begin{array}{c}\left (\vert 00 \rangle _{B} +\vert  01 \rangle _{B} +\vert 10 \rangle _{B} +\vert  11 \rangle _{B}\right ) \\
\left (\vert 00 \rangle _{A} +\vert 01 \rangle _{A} +\vert 10 \rangle _{A} +\vert 11 \rangle _{A}\right )\end{array} & \, & \, \\
\Downarrow  & \, & \, \\
\begin{array}{c}\left (\vert 00 \rangle _{B} +\vert  01 \rangle _{B}\right ) \\
\left (\vert 00 \rangle _{A} +\vert 01 \rangle _{A} +\vert 10 \rangle _{A} +\vert 11 \rangle _{A}\right )\end{array} &  \Rightarrow \hat{\mathbf{U}}_{1 ,2} \Rightarrow  & \vert 00 \rangle _{B}\vert  00 \rangle _{A} +\vert 01 \rangle _{B}\vert  01 \rangle _{A} \\
\, & \, & \Downarrow  \\
\vert 01 \rangle _{B}\left (\vert 00 \rangle _{A} +\vert 01 \rangle _{A} +\vert 10 \rangle _{A} +\vert 11 \rangle _{A}\right ) &  \Leftarrow \hat{\mathbf{U}}_{1 ,2}^{\dag } \Leftarrow  & \vert 01 \rangle _{B}\vert  01 \rangle _{A}\end{array}\tag{Table III}
\end{equation}

\medskip

 The initial measurement of $\hat{B}_{l}$, selecting the
$0$
of
$01$, projects the initial quantum superposition of all the basis vectors of register $B$ on the superposition of those
beginning with
$0$
-- vertical arrow on the left of the diagram.  The latter superposition causally propagates forward in time, by
$\hat{\mathbf{U}}_{1 ,2}$, into the superposition of the two tensor products on the right of the right looking horizontal arrows. Then the
final measurement of
$\hat{A_{r}}$, selecting the
$1$
of
$01$, projects the latter superposition on the term ending with $1$ under the right vertical arrow. Note that this outcome of the final measurement, which inherits both selections, coincides with that of the ordinary quantum description of the quantum algorithm -- see the bottom-right corner of Table I diagram. Time-symmetrically, this final measurement outcome causally propagates backward in time, by
$\hat{\mathbf{U}}_{1 ,2}^{\dag }$, until becoming the definitive outcome of the initial measurement. This backward in time propagation (which inherits both selections) is an instance of the time-symmetrization of the ordinary quantum description
of the quantum algorithm -- bottom line of the diagram. Note that, mathematically (disregarding the time-direction of causality) it is the ordinary quantum description of the unitary part of the quantum algorithm back again -- see the bottom line of the diagram in Table I. Indeed, the symbol ``$ \Leftarrow \hat{\mathbf{U}}_{1 ,2}^{\dag } \Leftarrow $'' is mathematically equivalent to ``$ \Rightarrow \hat{\mathbf{U}}_{1 ,2} \Rightarrow $''.

Also note that, together, the forward and backward in time propagations form a causal loop that changes its own past;  the outcome of its initial measurement of
$\hat{B}_{l}$  is changed into the outcome of the initial measurement of
$\hat{B}$  of the ordinary quantum description; compare the state of register $B$ under the left vertical arrow with the state of register $B$ in the bottom-left corner of the diagram.

We should eventually take the quantum superposition of all the time-symmetrization instances. We can see that, no matter how we evenly share the selection between initial and final measurements, the final measurement outcome, which inherits both selections, is always that of the final measurement of
$\hat{A}$  of the ordinary quantum description of the quantum algorithm. Correspondingly, its subsequent backward in time propagation by
$\hat{\mathbf{U}}_{1 ,2}^{\dag }$ (bottom line of the diagram), mathematically (disregarding the time-direction of causality), is always the ordinary quantum description of the unitary process between the two one-to-one correlated measurement outcomes, namely the bottom line of Table I diagram.

Summing up, mathematically, time-symmetrization leaves the ordinary quantum description of the unitary evolution between the two measurement outcomes unaltered but changes the structure of causality along this evolution. The single causal process of the ordinary quantum description is replaced by a quantum superposition of causal loops.

\subsubsection{
Relativizing the causal loops with respect to the problem-solver
}
The causal loops generated by time-symmetrizing the ordinary quantum description of the quantum algorithm, which is with respect to the external observer, do not yet  map on those logically implied by the sheer existence of the quantum speedup, which are with respect to the problem-solver. Indeed, we should relativize the causal loops of Table III with respect to the problem-solver. We should just conceal from her the part of the problem-setting selected by the initial measurement. This is done by postponing after her problem-solving action  the projection of the quantum state due to the initial measurement   of
$\hat{B}_{l}$ . This yields the causal loop in the following table:

\medskip

\begin{equation}\begin{array}{ccc}\begin{array}{c}\;\text{time }t_{1}\text{, meas. of}\;\hat{B}_{l}\end{array} & t_{1 \rightleftarrows }t_{2} & \text{time }t_{2}\text{,meas. of}\;\hat{A_{r}} \\
\, & \, & \, \\
\begin{array}{c}\left (\vert 00 \rangle _{B} +\vert  01 \rangle _{B} +\vert 10 \rangle _{B} +\vert  11 \rangle _{B}\right ) \\
\left (\vert 00 \rangle _{A} +\vert 01 \rangle _{A} +\vert 10 \rangle _{A} +\vert 11 \rangle _{A}\right )  \end{array} &  \Rightarrow \hat{\mathbf{U}}_{1 ,2} \Rightarrow  & \begin{array}{c}\vert 00 \rangle _{B}\vert  00 \rangle _{A} +\vert 01 \rangle _{B}\vert  01 \rangle _{A} + \\
\vert 10 \rangle _{B}\vert  10 \rangle _{A} +\vert 11 \rangle _{B}\vert  11 \rangle _{A}\end{array} \\
\, & \, & \Downarrow  \\
\begin{array}{c}\left (\vert 01 \rangle _{B} +\vert  11 \rangle _{B}\right ) \\
\left (\vert 00 \rangle _{A} +\vert 01 \rangle _{A} +\vert 10 \rangle _{A} +\vert 11 \rangle _{A}\right )\end{array} &  \Leftarrow \hat{\mathbf{U}}_{1 ,2}^{\dag } \Leftarrow  & \vert 01 \rangle _{B}\vert  01 \rangle _{A} +\vert 11 \rangle _{B}\vert  11 \rangle _{A}\end{array}\tag{Table IV}
\end{equation}

\medskip

 The projection of the initial state of register $B$ due to the measurement of
$\hat{B}_{l}$  -- top-left corner of the diagram -- is postponed after Alice's problem-solving action, namely outside the present table which is limited to that action. With this postponement, the initial state of register $B$ goes unaltered through the initial measurement of
$\hat{B}_{l}$  becoming, by $\hat{\mathbf{U}}_{1 ,2}$, the quantum superposition of tensor products above the vertical arrow. Alice's final measurement of
$\hat{A_{r}}$, which in this time-symmetrization instance selects the value $1$ of the right bit of register $A$, projects the state above the vertical arrow on the state below it. Then (in causal order), this final measurement  outcome propagates backward in time by $\hat{\mathbf{U}}_{1 ,2}^{\dag }$ until becoming the definitive outcome of the initial measurement. This backward in time propagation is an instance of the time-symmetrized description of the quantum algorithm with respect to the problem-solver. It tells her, before she begins her problem-solving action, one of the possible halves of the information about the solution.

Note that we would have obtained the identical diagram by time-symmetrizing the description of the quantum algorithm relativized to the problem-solver of Table II.

Also note that, for the mathematical equivalence between $ \Leftarrow \hat{\mathbf{U}}_{1 ,2}^{\dag } \Leftarrow $ and $ \Rightarrow \hat{\mathbf{U}}_{1 ,2} \Rightarrow $, the  bottom line of Table IV diagram (mathematically) can also be read in the usual left to right way:

\medskip

\begin{equation}\begin{array}{ccc}\text{time }t_{1} & t_{1} \rightarrow t_{2} & \text{time }t_{2} \\
\, & \, & \, \\
\begin{array}{c}(\vert 01 \rangle _{B} +\vert  11 \rangle _{B}) \\
\left (\vert 00 \rangle _{A} +\vert 01 \rangle _{A} +\vert 10 \rangle _{A} +\vert 11 \rangle _{A}\right )  \end{array} &  \Rightarrow \hat{\mathbf{U}}_{1 ,2} \Rightarrow  & \vert 01 \rangle _{B} \vert 01 \rangle _{A} +\vert  11 \rangle _{B}\vert 11 \rangle _{A}\end{array}\tag{Table V}
\end{equation}

\medskip

Summing up, the entire causal loop is the zigzag diagram of table IV. We can see that it is one of the causal loops whose mutually exclusive $or$ is logically implied by the sheer existence of the quantum speedup. The state of register $B$  in the top-left part of the diagram tells us that Alice, before beginning her problem-solving action, is completely ignorant of the number of the drawer with the ball. Through the zig-zag, this state changes into the state of register $B$ in the bottom-left part of the diagram, which tells us that Alice, before beginning her problem-solving action, knows in advance that the ball is in drawer either $01$ or $11$. She can then solve the problem in an optimal logical way by opening just either drawer. In equivalent terms, the computational complexity of the problem to be solved by Alice quadratically reduces from the problem of locating the ball in one of $4$ drawers to that of locating it in $2 =\sqrt{4}$ drawers. In still equivalent terms, the dimension of the Hilbert space in which the search for the solution takes place quadratically reduces from $4$ basis vectors to $2 =\sqrt{4}$ basis vectors.

\subsection{
Generalization
}
Until now we have focused on the simplest instance of Grover algorithm, but the results obtained go well beyond it.

First, everything immediately generalizes to any number, $N$, of drawers: it suffices to increase from $2 =\sqrt{4}$ to $\sqrt{N}$ the number of quantum bits in each quantum register. We can see that all considerations remain unaltered; Alice (the problem-solver) always knows in advance one of the possible halves of the information about the solution and can use this knowledge to produce the solution with fewer computation steps, $\ensuremath{\operatorname*{O}}\left (\sqrt{N}\right )$ in the optimal quantum case. In equivalent terms, the dimension of the  Hilbert space in which the search for the solution (or for one-to-one correlation between the two observables) takes place reduces from $N$ basis vectors to $\sqrt{N}$ basis vectors.

Second, we never needed to look inside
$\hat{\mathbf{U}}_{1 ,2}$. This unitary transformation is just defined by the fact that it must build up one-to-one correlation between two initially unrelated commuting observables ($\hat{B}$ and
$\hat{A}$) without changing the eigenvalue of the observable ($\hat{B}$) selected by the initial measurement, like in Table I in the case of $2$ quantum bits per register. Besides Grover algorithm, we could be dealing with any quantum algorithm where the solution is an invertible function of the problem-setting and is determined with certainty. In any such quantum algorithm, the problem-solver always knows in advance half of the information about the solution she will produce and measure in the future and can use this knowledge to produce the solution with fewer computation steps. In the optimal quantum case, the number of computation steps is that needed to solve the problem in an optimal logical way benefiting of the advanced knowledge in question. Of course, this explains the quadratic quantum speedup of Grover quantum search algorithm, where the search is in an unstructured domain. In $\left [4\right ]$, we have seen that, when the search is in a structured domain, the advanced knowledge of half of the information about the solution can create the room for an exponential speedup.

Things generalize also beyond quantum algorithms. Let us consider a unitary evolution between an initial and a final one-to-one correlated observables that does not change the eigenvalue of the observable selected by the initial measurement. In this kind of process, one is free to postpone the initial measurement at will. When made simultaneous with the final measurement, the time-symmetrization we are dealing with must already be in place to become the perfect atemporal symmetry between the simultaneous measurements of two maximally entangled observables -- see the states above and below the vertical arrow in Table II, thinking that the measurement of $\hat{B}$ is postponed and made simultaneous with that of
$\hat{A}$. This is for the principle of sufficient reason: there would be no reason for the prevalence of either measurement in the simultaneous selection of the pair of one-to-one correlated measurement outcomes.

Note that, in the case of quantum algorithms, the above reason for time-symmetrizing is redundant with the previous reason that that is necessary being the only way of generating the quantum superposition of the causal loops whose mutually exclusive $or$ is logically implied by the sheer existence of the quantum speedup. Since the time-symmetrization we are dealing with changes our way of viewing physical reality, the redundancy in question should not be despised.

For the more general reason, we can state that:

\medskip

\textit{The ordinary quantum description of any unitary evolution between an initial and a final one-to-one correlated measurement outcomes that does not change the eigenvalue selected by the initial measurement is incomplete and is completed by time-symmetrizing it. }

\medskip

 Let us call the above statement:\textit{ the causal loop rule of quantum correlation. }The cases are two: 

\medskip

(i) The unitary evolution builds up one-to-one correlation between two initially unrelated observables. This is, of course, the case of quantum algorithms but it can also be that any process of identical form. We mean the form of Table I in the case of the two quantum bits per register. Note that all the considerations of the previous subsection apply and the building up of the quantum correlation can always occur with an at least quadratic quantum speedup. Of course, if the process is not a quantum algorithm, the notion of advanced knowledge, on the part of the problem-solver, of half of the information about the solution she will produce and measure in the future must be replaced by the notion of an equivalent reduction of the dimension of the Hilbert space (from $N$ basis vectors to $\sqrt{N}$ basis vectors) in which the search for one-to-one correlation takes place.

\medskip

(ii) The unitary evolution does not build up any quantum correlation because the two observables are maximally entangled to start with. We have seen in the Rationale that this is the case of quantum nonlocality. Here, the causal loops generated by the time-symmetrization play the role of the hidden variables envisaged by Einstein and the others to locally explain quantum nonlocality.

\medskip

We should eventually note that the causal loops we are dealing with should be exempt from the paradoxes that plague the usual kind. In fact, the time-symmetrization that generates them is imposed by a quantum mechanical symmetry; therefore, as long as quantum mechanics is consistent, they should be exempt from any paradox. When needed to distinguish them from the usual kind, we will call them \textit{time-symmetric causal loops} (being generated by a time-symmetrization procedure).

Let us conclude this part of the work by mentioning priorities along the path that led to the discovery of the quantum computational speedup, which, according to the present work, would be an even more fundamental discovery than currently thought. In fact, it would introduce a special form of retrocausality (the one involved in the time-symmetric causal loops) in the non-relativistic quantum world.

 The notion that computation is possible in the quantum world and the fundamental notion of quantum bit was introduced by Finkelstein $[20]$ in 1969; that was in the context of seeing the universe as a quantum computer (a cellular automaton), while searching for a unification between quantum mechanics and relativity theory. In 1982, Feynman $\left [21\right ]$ highlighted the higher than classical computational efficiency of the quantum world, noticing that the classical simulation of a quantum process can require an amount of physical resources exponentially higher than the amount involved in the quantum process itself. In 1985, Deutsch $\left [22\right ]$ devised the seminal quantum algorithm that delivers a quantum computational speedup.

\section{
The possible physical basis of the teleological evolutions of life
}
The fact that quantum algorithms are full fledged teleological evolutions raises the question whether the physical basis of their teleological character could also apply to the natural evolutions to which the teleological notion was once applied -- is sometimes applied also today.

We have seen in the Rationale that the only possible way of seeing quantum features in the evolutions of life, which can be the size of Earth and have durations of billions of years, is deriving them from an assumed quantum evolution of the universe. 

A consequent quantum interpretation of the Fine-tuned Universe version of the Anthropic Principle says that the evolution of the universe toward the ``value'' of life (in particular, ``existence of sentient life'' or ``absence of life'') under the fundamental physical laws and constants builds up a one-to-one quantum correlation between the value of the fundamental constants since the beginning of time and the value of life today. The evolution in question can therefore occur with quantum speedup and consequently be teleological in character. The ``can occur'' is in the sense that there is certainly the room for its occurring with quantum speedup. Like in the case of random quantum search algorithms $\left [15\right ]$, under the (conservative) assumption that the search for this one-to-one correlation is random, the room for quantum speedup becomes an actual quantum speedup. In the universe we are in, of course, the same would apply to the evolution of the universe toward life and, cascading, to the decohered evolution of life on Earth and the evolutions of species.

In the following subsections, after simply encoding a toy model of the evolution of the universe toward the value of life on the quantum description of Grover quantum search algorithm, we examine the impact that a teleological character of the evolution of the living would have on different areas of science and knowledge.

\subsection{Teleological evolution of the universe toward the value of life -- a toy model 
}
A toy model of a teleological evolution of the universe toward the value of life can trivially be mapped onto the quantum description of Grover algorithm.

We should imagine a \textit{value of the fundamental constants} observable
$\hat{B}$  and a\textit{ value of life} observable
$\hat{A}$. Say that the eigenvalue $01$ of
$\hat{B}$  is the value of the fundamental constants compatible with the development of sentient life and the eigenvalues $00 ,10 ,11$ the values of those incompatible. Correspondingly, let the eigenvalue $01$ of
$\hat{A}$ mean existence of sentient life; the eigenvalues $00 ,10 ,11$ absence of life.

Initially,
$\hat{B}$ should be in a quantum superposition (or, indifferently, mixture) of all its eigenstates (i.e of all the possible universes each with its value of the fundamental constants). We can assume that also
$\hat{A}$, the value of life observable, is initially in an independent quantum superposition of all its eigenstates.

The universe's unitary evolution under the fundamental physical laws and constants would have finally, at the present time, built up a quantum superposition of tensor products, each an eigenstate of
$\hat{B}$ multiplying the corresponding eigenstate of
$\hat{A}$ -- see Table II.  In the universe we are in, at the present time one with observers, the measurement of            $\hat{A}$ at the present time would select the eigenvalue $01$ of
$\hat{A}$, namely ``existence of sentient life'' and the corresponding eigenvalue $01$ of the value of the fundamental physical constants observable
$\hat{B}$. Since the universe's evolution does not change the value of the fundamental constants, it is as if the eigenvalue of
$\hat{B}$  was selected by a measurement of
$\hat{B}$ at the beginning of time, when there could be no observers.

Of course, this is exactly the quantum search algorithm described in the previous section, the version with respect to the problem-solver of Table II. As shown by its time-symmetrization (Table IV), it can undergo with a quantum speedup (no matter the specific form of
$\hat{\mathbf{U}}_{1 ,2}$) and does so even if the quantum search for the solution is random $\left [15\right ]$.

\subsection{
Decohered teleological evolutions: what would they look like? 
}

\medskip  Decohered and endowed with quantum speedup, it is natural to wonder what the supposedly teleological evolutions of life would look like. All natural evolutions are necessarily described in a probabilistic way. For example, the stochastic description of a Darwinian evolution can be an alternation of random changes in the species DNA pool each followed by natural selection of the DNA changes more suited to the environment. Of course, in a classical description, the transition probabilities can depend on the past of the process, not on the future of it. Instead, a teleological process must have an attractor in its future. This means that those DNA mutations that in the future will survive natural selection because more fit to the environment should have the transition toward them more probable than in the classical case.

Let us mention what could be an emblematic example of this kind of natural evolution:

 It is known that, for many millions of years, the ancestors of the flying birds were feathered dinosaurs who did not fly. Very light, strong but flexible, cave, and air-proof, feathers are of course ideal for flight. So, why anticipating something designed for flight many millions of years in advance? 

The usual explanation is of course a causal one: at first, feathers appeared by random mutation and survived natural selection as good thermal insulators, later they proved useful for speeding escape from predators, perhaps for gliding from trees to the ground, and only eventually they turned out to be what was needed to conquer the air. 

In the teleological view, instead, even during the flightless period of the  feathered dinosaurs those random DNA mutations that would have led in the future to the conquest of the air would have more likely occurred. 

 Of course, this teleological explanation of the evolution of feathers is puzzling. However, let us note that it is no more puzzling than the causal loop explanation of the quantum computational speedup. We are in the presence of a quantum cosmological assumption that can make the evolutions of life as puzzling as the phenomena of the microscopic quantum world.

\subsection{
How to  model teleological natural evolutions
}
We address the problem of how to model a decohered evolution endowed with quantum speedup -- as that of the living on Earth should be. Of course, the canonical way would be to obtain this evolution from that of the universe toward life by tracing out all the rest. Although this, in the present case, obviously holds only in principle, it would be interesting to investigate toy models of such an evolution.

One should identify the purely quantum evolution of a system (standing for the universe) toward a goal (life), endowed with quantum speedup, where the achievement of the goal is mostly localized in a subsystem thereof (standing for the Earth). Then, one should obtain the decohered evolution of the subsystem from that of the system by tracing out all the rest. Note that the interaction of the subsystem with the rest of the system, in the present case, would do the opposite than decoherence. It would be what allows the evolution of the subsystem toward the goal to occur with quantum speedup as part of the purely quantum evolution of the system.

Let us note that the evolution of life on Earth should have been almost completely independent of the evolution of the rest of the universe. If it were completely independent, under the quantum cosmological assumption it would be factorizable out of the evolution of the universe, thus itself quantum and endowed with quantum speedup. Therefore, an obvious, approximate, way of modeling the decohered evolution of the living on Earth endowed with quantum speedup is just to ignore the fact that it is decohered and model it in a quantum way. There are encouraging precedents to this. For example, it has been shown that modeling in a quantum way the interplay between species of predators competing for the same preys allows to explain the empirically ascertained species' asymptotic behavior in a way otherwise difficult to explain $\left [23\right ]$.

A further way of modeling a decohered evolution endowed with quantum speedup should be to model it as a stochastic process with transition probabilities depending on both the past and the future of the process, something just outlined for the time being.

\subsection{
Impact on the natural sciences
}
The current position of the natural sciences toward teleology can be considered in some measure conflicting. On the one hand, the teleological view would seem to allow a deeper understanding of the functional structures and the evolutions of the living. On the other, as this vision is considered without any possible physical basis and therefore misleading, there is outright hostility to it.

This tension is captured by the following sentence of the evolutionary biologist Haldane $\left [24\right ]$: 

\medskip

\textit{Teleology is like a mistress to a biologist: he cannot live without her, but he is unwilling to be seen with her in public}.

\medskip

It would seem that, today, only a few scientists support the teleological view. An example is the biologist Youvan $\left [25\right ]$, who shows the revolutionary character that a retrocausal physics would have on the theory of cognition. Another example is that of the philosopher Nagel $\left [26\right ]$, who argues that the principles that explain the emergence of life should be teleological rather than materialistic or mechanistic in character.

Of course, the sheer notion that quantum computation is a scientific example of teleological evolution may already put a strain on the present position of biology.

In recent years, a group of evolutionary biologists $\left [27\right ]$ expressed the need to rethink Darwin's theory of evolution. That was after the discovery that some species' evolution steps occurred far too quickly to be solely ascribed to random changes in the species' DNA pool. Of course, the present teleological view of evolutions, with the associated quantum speedup, would offer a possible explanation.

As already mentioned, the teleological view would also go along with the fact that some physicists have began to apply essentially quantum mechanical features like quantum entanglement and nonlocality to the macroscopic game of species, showing that this can explain some empirically ascertained features of it otherwise difficult to explain $\left [23\right ]$.

\subsection{
Impact on philosophy
}
Calling philosophy into play is unconventional in a work of physics. However, we are of course in a very particular case. Indeed, the present quantum theory of teleological evolutions turns out to be the formalization of a  notion of natural philosophy dating back two and a half millennia. $\left [28\right ]$. We should thus be entitled to look for other possible cases where philosophical thinking has anticipated aspects of the present scientific notion.

In effect, the teleological theory in question seems to revamp a full body of philosophical views sort of opposite to the body of philosophical views emerged with the scientific revolution.

For short, let us call the present teleological view of evolutions and the causal view of today's physics and the natural sciences respectively the\textit{ mutual causality view}\protect\footnote{
We can say that what generates the quantum superposition of the time-symmetric causal loops is the \textit{mutual causality} between the initial and final measurement outcomes. 
} and the \textit{causal view}. The causal view would underlie the materialist, mechanistic and reductionist views, while the mutual causality view would underlie the idealist, teleological and monist views.

Consider the opposition between materialism and idealism. The former, about the observer being causally created by physical reality, is a causal view; the latter, about the mutual causation between observer and physical reality, like in the participatory Anthropic Principle of Wheeler $\left [29\right ]$, and the anthropic cosmological principles of  Barrow and Tipler $\left [8\right ]$ is a mutual causality view. 

For what concerns the opposition between the mechanistic and teleological views of evolutions, it is obviously the opposition between the causal and mutual causality views of evolutions.

Regarding reductionism vs. monism, note that deriving the teleological character of the decohered evolutions of the living on Earth from that of an assumed quantum evolution of the universe toward life is an essentially monistic view. It presupposes that the description of what we see around and within us is essentially inseparable from that of the whole.

Let us mention other theories in the history of philosophy reminiscent of the mutual causality between sentient life today and the physical laws with their physical constants since the beginning of time:\

\medskip

(i) For Plato, ``Ideas [of our mind] are objective perfections that exist in themselves and for themselves and, at the same time, they are the causes of natural phenomena and constitute their unity'' $\left [30\right ]$; in modern language, the cause of natural phenomena that constitute their unity are the fundamental physical laws. In the present teleological view, this might correspond to the identity and mutual causality between the fundamental physical laws (and their fundamental constants) since the beginning of time and the ideas of our mind. This should also go along with Fritjov Capra's $\left [31\right ]$ notion that there is identity between the fundamental states of consciousness described by eastern theosophies and the fundamental laws of quantum physics. By the way, we do not need to be anthropocentric. The above would also apply to spatial intelligence, often superior to ours in non-human animals.

Also Plato's notion that ideas -- fundamental laws -- are objective perfections could have a correspondence with the Fine-tuned Universe version of the Anthropic Principle, with the fact it asserts that the slightest change of the value of the fundamental constants would have produced a radically different universe most probably unable to host life.

\medskip

(ii) The present notion of teleological evolution, namely of an evolution toward a goal with an attractor in the goal it will produce in the future, could also be seen in terms of a ``systematically lucky'' evolution, lucky in reaching that goal because attracted by it. Interestingly, also the notion of a systematically lucky evolution seems to have a precedent in pre-scientific thinking. It could be the Stoic notion of $\pi \rho o\nu o\iota \alpha $ (foreknowledge)\textit{$\left [32\right ]$}, sometimes also translated into \textit{providence} and, in the present context, possibly translatable into \textit{advanced knowledge} (e. g., of half of the information about the solution of a problem). Foreknowledge, or advanced knowledge, would have pushed the evolution toward its goal in a luckier and thus faster way than what is possible in the case of a causal evolution, exactly as the advanced knowledge about the solution does with quantum algorithms. This notion of $\pi \rho o\nu o\iota \alpha $ would also fit well with the notion of teleological evolution understood as a stochastic processes with transition probabilities time-symmetrically affected by the past and future of the process, with the luckier transitions more probable than in the classical case.

\medskip

(iii) Michelangelo's ``Creation of Adam'' on the ceiling of the Sistine Chapel could be the \textit{manifesto} of the present theory. There is a mirror symmetry between creator and created $\left [33\right ]$. Could this be interpreted as the time-symmetry -- and the related mutual causation -- between the fundamental physical laws with their physical constants since the beginning of time and the existence of sentient life today?

\section{
Research prospects
}
The teleological view, with a \textit{present} time-symmetrically affected by the \textit{past} and the \textit{future}, changes the very way of viewing physical reality. Correspondingly, research prospects on the subject could have a broad spectrum.

\medskip For what concerns quantum physics: 

\medskip

(i) Look for a possible synthesis between the present notion of the time-symmetric causal loops and time-symmetric quantum mechanics $\left [34 -38\right ]$ to which it is obviously inspired.

\medskip

(ii) The advanced knowledge rule provides the number of oracle queries needed to solve in an optimal quantum way any oracle problem. The unifying character of this rule might suggest the possibility that there is a universal optimal quantum algorithm for this class of problems. Looking for this possibility could be a research prospect.

\medskip

\medskip
(iii) Look for toy models that capture the mechanism for which the mostly autonomous evolution of a subsystem (the evolution of life on Earth), although decohered, inherits the quantum speedup and the consequent teleological character of the evolution of the system (the evolution of a quantum universe toward life).

\medskip

(iv) Look for a possible impact of the teleological view on quantum cosmology. According to current assumptions, also the ``inanimate'' part of the evolution of the universe should be an evolution toward life and would therefore have an attractor in its present state. Incidentally, since the assumption that the universe evolves in a unitary quantum way is essential to make a teleological character of natural evolutions possible, a possible empirical evidence of the teleological character of the evolutions of the living would also be empirical evidence of the quantum cosmological assumption.

\medskip

(v) Look for toy models of teleological Darwinian evolutions. Note that these evolutions could be essentially different from quantum algorithms. For example, the fact that their teleological character should also be reflected in the functional structure of the living would have no parallel with quantum algorithms.

\medskip

 (vi) Look for applications of the teleological view to quantum self-organization. For example, one could look for toy models of the leap from a soup of life precursor molecules to a self-replicating system, a possibility considered in abiogenesis. Seeing this awkward (classical) leap as the part of a unitary quantum evolution with an attractor in a future full fledged self-replicating system might be appealing.

\medskip

 (vii) Finally, opposing schools of thought throughout the history of philosophy could be reviewed in the light of the opposition between the causal and mutual causality views, as exemplified in the previous subsection. The fact that the philosophical notion of teleological evolution has a physical basis might reduce the distance between the physical and philosophical views; there might be further room for cross-fertilization.

\medskip

The above prospects are imagineable extensions of the present work. However, the notion that the quantum framework can host a special form of retrocausality might have a broader impact on the physical view of reality.

\section{
Positioning the work
}

Of course, we are not alone in challenging the predominant idea that causality always goes in the same (forward) time-direction. Other approaches to quantum mechanics that share this feature are time-symmetric quantum mechanics $\left [34 -38\right ]$, particularly the study of the possibility that a future choice changes a past quantum measurement outcome $\left [37\right ]$, and ongoing attempts to study quantum processes without the need to assume a definite causal structure $\left [39 -43\right ]$. 

The present retrocausal approach to the quantum speedup is naturally inspired by time-symmetric quantum mechanics and, of course, has in common with the attempts in question the fact of looking beyond the ordinary notion of causality, not in common the approach. In fact, these attempts explicitly rule out the possibility that quantum algorithms, with their causally ordered quantum circuits, can accommodate causality violations $\left [40 ,41\right ]$. For us instead, with their quantum speedup over what is classically possible, they are the paradigmatic example of causality violation. The two approaches seem to address different kinds of causality violations.

Some concepts of the work have precedents.

One is the celebrated EPR paper $\left [7\right ]$. In it, Einstein, Podolsky, and Rosen already argued that the quantum description of quantum nonlocality is incomplete. As we have seen, the causal loops generated by time-symmetrizing the ordinary quantum description of quantum nonlocality play the role of the hidden variables they envisaged to physically (locally) explain it $\left [5\right ]$.

Another precedent is the local, retrocausal, explanation of quantum nonlocality given by Costa de Beauregard in 1953 $\left [44\right ]$. Consider the measurements in the same basis of two maximally entangled observables in a state of spatial separation of the respective subsystems. Costa de Beauregard assumed that the outcome of the first of the two measurements locally propagates backward in time, by the inverse of the unitary transformation that spatially separates the two subsystems, until these are not yet spatially separate. There, always locally, it would change the state of the other observable in a way that ensures the appropriate correlation between the two future measurement outcomes. 

This retrocausal explanation of quantum nonlocality is similar to ours as it is based on a causal loop. The difference is in the character of the loop. Sending backward in time an entire measurement outcome can originate paradoxes and, of course, is not imposed by quantum mechanics but is instrumental to finding a local explanation of quantum nonlocality. The time-symmetric causal loops originated by time-symmetrizing the ordinary quantum description are instead imposed by quantum mechanics. Therefore, as long as the latter is consistent, they should be exempt from the paradoxes that plague the usual kind.

Interestingly, at the time, Costa de Beauregard was an assistant of De Broglie's, who forbade him to publish that work for years. This, until Wheeler and Feynman published their emitter-absorber theory, which - for the first time in the history of physics - made use of retrocausality - see $\left [45\right ]$ for an account of this. This tells us a lot about the position of physicists, even at the top level, towards retrocausality. By the way, despite the importance of the author's name, Costa de Beauregard's explanation of quantum nonlocality never took off.

Still other precedents are Wheeler's Participatory Anthropic Principle $\left [29 ,46\right ]$ and the anthropic cosmological principle of Barrow and Tipler $\left [8\right ]$. These two principles share with the present work the notion of mutual causality between the observation today and the ``value'' of the fundamental physical laws and constants since the beginning of time mediated by a unitary evolution of the universe in between. Dating to before Deutsch's 1985 discovery of the quantum speedup, of course the two principles in question make no reference to it.

\section{Conclusions
}

We provide the line of thought followed in the work.

We have shown that the sheer existence of a quantum computational speedup logically implies the mutually exclusive $or$ of well defined causal loops. In each of them, it is as if the problem-solver knew in advance, before beginning her problem-solving action, one of the possible halves of the information that specifies the solution of the problem she will produce and measure in the future. She can then use this knowledge to produce the solution with fewer computation steps than classically possible. The number of computation steps needed to solve the problem in an optimal quantum way is that needed to solve it in an optimal logical way benefiting of the advanced knowledge in question.

Of course, these causal loops must be implicit in the quantum algorithm that delivers the quantum speedup. Since the ordinary quantum description of quantum algorithm does not and, being causal in character, cannot describe any causal loop, we must conclude that it is incomplete. We have seen that the only possible way of completing it is by time-symmetrizing it.

This cannot be done on the usual quantum description of quantum algorithms -- a unitary computation of the solution followed by its measurement -- which is itself trivially incomplete lacking the initial measurement. First, we must extend it to the upstream process of setting the problem. An initial measurement selects the problem-setting out of a quantum superposition (or mixture) of all the possible problem-settings; the unitary computation and measurement of the corresponding solution follow. We call this extended quantum description \textit{the ordinary quantum description of the quantum algorithm}, ordinary because is still a causal description that cannot describe causal loops.

Then, we time-symmetrize this ordinary quantum description by requiring that the determination of the problem-setting, and thus also that of the corresponding solution, evenly shares between the initial measurement of the problem-setting and the final measurement of the solution. For each possible way of evenly sharing, the part of the selection ascribed to the initial measurement unitarily propagates forward in time until becoming the state immediately before the final measurement; the part performed by the final measurement propagates backward in time, by the inverse of the time-forward unitary transformation, until replacing the previous outcome of the initial measurement. We should then take the quantum superposition of all these time-symmetrization instances. 

This time-symmetrization leaves the unitary part of the ordinary quantum description of the quantum algorithm mathematically unaltered but changes the description of the behavior of causality along it. The single causal process of the ordinary quantum description is replaced by a quantum superposition of causal loops; as seen by the problem-solver (who should be concealed from the part of problem-setting selected by the initial measurement), these are the causal loops whose mutually exclusive $or$ is logically implied by the sheer existence of the quantum speedup.

Since the time-symmetrization we are dealing with turns out to be the only possible way of obtaining a quantum description of the quantum algorithm that describes the causal loops that it must host, it is mandatory.

This quantum causal loop description of quantum algorithms immediately generalizes to a most general form of quantum correlation, that is, any unitary evolution between two one-to-one correlated measurement outcomes that does not change the eigenvalue selected by the initial measurement (the problem-setting in quantum algorithms). The time-symmetrization of the ordinary quantum description of this kind of process is imposed by the fact the initial measurement can be postponed at will. When made simultaneous with the final measurement, the time-symmetrization in question must already be in place to satisfy the symmetry implicit in the simultaneous measurements of two maximally entangled observables.

In particular, this generalization covers the case of quantum nonlocality. The causal loops generated by time-symmetrizing its ordinary quantum description play the role of the hidden variables envisaged by Einstein and the others to physically (locally) explain it. Thus, the incompleteness we are currently ascribing to the ordinary quantum description of quantum algorithm is exactly the same that Einstein and the others ascribed to the quantum description of quantum nonlocality. In both cases it is the incompleteness of the quantum description of a quantum correlation.

Eventually, the causal loop description of quantum algorithms highlights their teleological character, that is, their being evolutions toward a goal (the solution) with an attractor in the goal they will produce in the future (the solution again). 

The above naturally raises the question whether the quantum superposition of causal loops that stands at the basis of the quantum speedup could also be the physical basis of the teleological character of the evolutions of the living for which the notion of teleological evolution was conceived.

As we have seen, a quantum cosmological interpretation of the Fine-tuned Universe version of the Anthropic Principle shows that the evolution of the universe under the fundamental physical laws and constants builds up a quantum correlation between the value of the fundamental constants since the beginning of time and the ``value'' of life today (in particular, ``existence of sentient life'' or ``absence of life''). It can correspondingly be endowed with quantum speedup and thus be teleological in character. In the universe we are in, this possibility would be inherited by: (i) the evolution of the universe toward life, (ii) even if decohered, the evolution of life on Earth, and (iii) the single evolutions on Earth, as those of species. Under the conservative assumption that the search for the value of life by the evolution of the universe under the fundamental physical law and constants is random, the possibility of quantum speedup becomes an actual quantum speedup.

We would conclude as follows: 

Explanations of physical phenomena involving retrocausality have been proposed in the past in particular by Costa de Beauregard $\left [44\right ]$, Wheeler and Feynman $\left [10\right ]$, Aharonov, Cohen, and Elitzur $\left [37\right ]$, and Cramer $\left [48\right ]$. Intriguing as they are, it can be said that these explanations never took off, as no form of retrocausality ever entered mainstream quantum mechanics. 

The present work makes a similar attempt in the new context of the quantum computational speedup. This makes two significant differences:

\medskip

(i) One is the fact is that any quantum speedup with respect to what is classically possible logically implies the mutual exclusive $or$ of well defined causal loops whose quantum superposition must be implicit in the quantum algorithm that delivers it. This also highlights the teleological character of quantum algorithms, their being evolutions toward a goal (the solution) with an attractor in the very goal they will produce in the future (again the solution).

\medskip

(ii) The other is that the retrocausality implicit in the teleological notion, mathematically, is always replaceable by the interplay between quantum state superpositions and quantum interferences that characterize unitary evolutions, indeed the ordinary quantum description of the quantum algorithms. This shouldmake the notion of retrocausality much more palateable.

\medskip

Under the quantum cosmological assumption that the universe evolves in a unitary quantum way, the same causal loops that highlight the teleological character of quantum algorithms could be responsible for the teleological character of natural evolutions. Quantum cosmology is of course hypothetical; in any case, the assertion that a teleological character of the evolutions of the living has no possible physical basis would be falsified.

\section*{
Acknowledgments
}
Thanks are due to Eliahu Cohen, Artur Ekert, Avshalom Elitzur and David Finkelstein for useful comments throughout the whole of the present retrocausal approach to the quantum speedup, to Daniel Sheehan for organizing the San Diego AAAS Pacific Division conferences on retrocausality, a far-looking forum for the discussion of frontier, also unorthodox, physics, and to Mario Rasetti for organizing the first series of international workshops on quantum communication and computation, the Elsag Bailey-ISI Turin Villa Gualino workshops of the years 1991-96 that brought to the discovery of the major quantum algorithms.

\section*{
References
}
$\left [1\right ]$ Castagnoli, G. and Finkelstein D. R.: Theory of the quantum speedup. \textit{Proc. Roy. Soc. Lon.}, 457, 1799-1807 (2001)

$\left [2\right ]$ Castagnoli, G.: The quantum correlation between the selection of
the problem and that of the solution sheds light on the mechanism of the quantum speed up.
\textit{Phys. Rev. A} 82, 052334 (2010)

$\left [3\right ]$ Castagnoli, G.: Completing the Physical Representation of Quantum
Algorithms Provides a Quantitative Explanation of Their Computational Speedup. Found. Phys. 48, 333-354
(2018) 

$\left [4\right ]$
Castagnoli, G.,  Cohen, E., Ekert, A. K., and  Elitzur, A. C.. A Relational
Time-Symmetric Framework for Analyzing the Quantum Computational Speedup. \textit{Found Phys.}, 49, 10 (2019),            1200-1230.

$\left [5\right ]$ Castagnoli, G.: Unobservable causal loops as a way to explain both the quantum computational speedup and quantum nonlocality. Phys. Rev. A, 104, 032203 (2021)

$\left [6\right ]$ Grover, L. K.: A fast quantum mechanical algorithm for database
search. \textit{Proc. 28th Annual ACM Symposium on the Theory
of Computing}. ACM press New York, 212-219 (1996)

$\left [7\right ]$
Einstein, A., Podolsky, B., and Rosen, N.: Can Quantum-Mechanical Description of
Physical Reality Be Considered Complete? Phys. Rev. vol. 47, n. 777 (1935)

$\left [8\right ]$ Barrow, J. D. and Tipler, F. J.: The Anthropic Cosmological Principle. 1st edition 1986 (revised 1988), Oxford University Press (1988)

$\left [9\text{]}\right .$ Long, G. L.: Grover algorithm with zero theoretical failure rate.
Phys. Rev. A 64, 022307-022314 (2001) 

$\left [10\right ]$ Wheeler, J. A. and Feynman R. P.: Classical Electrodynamics in Terms of Direct Interparticle Action. Rev. Mod. Phys. 21, 425-433 (1949)

$\left [11\right ]$ Smolin, L., Susskind, L.: Smolin vs. Susskind : The Anthropic Principle. Edge Institute (2004)

$\left [12\right ]$ Wilson, D. S. and Sober, E.: Adaptation and Natural Selection Revisited. Journal of Evolutionary Biology, 24, 462-468 (2011)

$\left [13\right ]$ Tegmark, M.: My Quest for the Ultimate Nature of Reality. ISBN 0307599809n (2014)

$\left [14\right ]$ Ayala, F. J.: Evolution, Explanation, Ethics and Aesthetics: Towards a Philosophy of Biology. ISBN 9780128036938, Academic Press (2016)

$\left [15\right ]$
Shenvi, N.,Kempe, J., Whaley, B.: A Quantum Random Walk Search Algorithm. ArXiv:quant-ph210064 (2002)

$\left [16\right ]$ Rovelli, C.: Relational Quantum Mechanics. \textit{Int. J. Theor. Phys.} 35, 637-658
(1996)

$\left [17\right ]$ Fuchs, C. A.. On Participatory Realism. arXiv:1601.043603 quant-ph] (2016)

$\left [18\right ]$ Bell, J. S.: On the Einstein Podolsky Rosen Paradox. Physics Vol. 1, No. 3, 195-290 (1964)

$\left [19\right ]$ Ekert, A. K.: Quantum cryptography based on Bell's theorem. Phys. Rev. Lett. 67 (6), 661-663 (1991)

$\left [20\right ]$ Finkelstein, D. R.: Space-Time Sructure in High Energy Interactions. In Fundamental Interactions at High Energy, Gudehus, T., Kaiser, and G., A. Perlmutter G. A. editors. Gordon and Breach, New York 324-338 (1969). Online PDF at http://homepages.math.uic.edu/\symbol{126}kauffman/FinkQuant.pdf

$\left [21\right ]$  Feynmann, R. P.: Simulating physics with computers. Int. J. Theor. Phys. 21, 467-488 (1982)

$\left [22\right ]$
Deutsch D.: Quantum theory, the Church Turing principle and the
universal quantum computer. Proc. Roy. Soc. A 400, 97-117(1985).

$\left [23\right ]$ Peled, B., Te'eni, A., Cohen, E., Carmi, A.: Study of entanglement via a multi-agent dynamical quantum game. PLoS ONE 18(1): e0280798. https://doi.org/10.1371/journal.pone.0280798 (2021)

$\left [24\right ]$ Dronamraju, K. R.: Haldane and Modern Biology. Johns Hopkins University Press, Baltimore, Maryland  (1968)

$\left [25\right ]$ Youvan, D. C.: Retrocausal Physics
and Human Cognition: Rethinking Consciousness, Logic and Thought. Preprint, DOI 10.13140/RG.2.2.18951.61609 (2024)

$\left [26\right ]$
Nagel, T.: Mind and Cosmos: why the materialist, neo-Darwinian conception of nature is almost certainly false. Oxford New York, Oxford University Press, ISBN 9780199919758 (2012)

$\left [27\right ]$
Laland, K., Uller, T., Feldman, M. et al. Does evolutionary theory need a rethink? Nature 514, 161-164 (2014)

$\left [28\right ]$ Lennox, J. G.: Aristotle Philosophy of Biology: Studies in the Origin of Life Science. Cambridge University Press (2000)

$\left [29\right ]$ Deutsch, D.: On Wheeler's notion of ``law without law'' in physics. Foundations of Physics, vol. 16, 565-572 (1986)

$\left [30\right ]$ Abbagnano, N.: STORIA DELLA FILOSOFIA. Torino-UTET, Vol. I, p 87 (1946)

$\left [31\right ]$ Capra, F.: The Tao of Physics: An Exploration of the Parallels Between Modern Physics and Eastern Mysticism. Shambhala Publications, Inc., 35th Anniversary Edition (2010)

 $\left [32\right ]$
Jacques Brunschwig, J.: Stoic Metaphysics. In The Cambridge Companion to Stoics, ed. B. Inwood, Cambridge, 206-232 (2006)

$\left [33\right ]$ Finkelstein, D. R.: Private communication. His noticing the mirror symmetry in question likely inspired the present work.

$\left [34\right ]$ Aharonov, Y., Bergman, P. G., and Lebowitz, J. L.: Time Symmetry
in the Quantum Process of Measurement. \textit{Phys. Rev.} 134,
1410-1416 (1964)

$\left [35\right ]$ Dolev, S. and Elitzur, A. C.: Non-sequential behavior of the wave
function. arXiv:quant-ph/0102109 v1 (2001)

$\left [36\right ]$ Aharonov, Y. and Vaidman, L.: The Two-State Vector Formalism: An
Updated Review. \textit{Lect. Notes Phys.} 734, 399-447 (2008)

$\left [37\right ]$ Aharonov, Y., Cohen, E., and Elitzur, A. C.: Can a future choice
affect a past measurement outcome? \textit{Ann.
Phys.} 355, 258-268 (2015)

$\left [38\right ]$ Aharonov, Y., Cohen E., and Tollaksen, J.: Completely top-down
hierarchical structure in quantum mechanics. \textit{Proc. Natl. Acad. Sci. USA, }115, 11730-11735
(2018)

$\left [39\right ]$
Oreshkov, O., Costa, F. \& Brukner, \v{C}. Quantum correlations with no causal order. Nat. Commun. 3, 1092 (2012).

$\left [40\right ]$ Ara{\'u}jo, M. et al. Witnessing causal nonseparability. New J. Phys. 17, 102001 (2015).

$\left [41\right ]$ Branciard, C. Witnessing causal nonseparability: an introduction and a few case studies. Sci Rep. 266018. https://doi.org/10.1038/srep26018, (2016)

$\left [42\right ]$ Milz, S., Pollock, F. A., Thao P Le, Chiribella, G. and Modi, K.: Entanglement, non-Markovianity, and causal non-separability. New Journal of Physics, vol. 20 (2018)

$\left [43\right ]$ Barrett, J., Lorenz, R.
and Oreshkov, O.: Cyclic quantum causal models. Nature Communications 12, 885 (2021)

$\left [44\right ]$
Costa de Beauregard, O. La m{\'e}canique quantique. Comptes Rendus Academie des Sciences , 236, 1632-34  (1953)

$\left [45\right ]$  Price, H. and Wharton, K.:  Disentangling the Quantum World.  arXiv:1508.01140 (2015)

$\left [46\right ]$
Thorne, K. S. and Zurek, W. H.: John Archibald Wheeler: A few highlights on his contributions to physics, Found. Phys. 16, 123-133 (1986)

$\left [47\right ]$ Cramer J. G.: The Quantum Handshake: Entanglement, Nonlocality and Transactions. Springer Verlag, chapter 7 (2015)

\end{document}